\documentclass[%
 aip,
 amsmath,amssymb,
 reprint,%
]{revtex4-1}

\usepackage{xr}
\usepackage{float}
\usepackage{cancel}

\usepackage{soul}
\usepackage{color}
\newcommand{\degree}{$^\circ$}

\usepackage{graphicx}
\usepackage{dcolumn}
\usepackage{bm}

\usepackage[utf8]{inputenc}
\usepackage[T1]{fontenc}
\usepackage{mathptmx}
\usepackage{etoolbox}
\usepackage{textcomp}

\usepackage{mathrsfs} 

\usepackage{xcolor}

\makeatletter
\def\@email#1#2{%
 \endgroup
 \patchcmd{\titleblock@produce}
  {\frontmatter@RRAPformat}
  {\frontmatter@RRAPformat{\produce@RRAP{*#1\href{mailto:#2}{#2}}}\frontmatter@RRAPformat}
  {}{}
}%
\makeatother
\begin{document}

\preprint{AIP/123-QED}

\title[A relationship between charge 
and work function measurement in SKPM]{
    A duality between surface charge and work function in
    scanning Kelvin probe microscopy}
\author{Isaac C.D. Lenton*}
 \email{isaac.lenton@manchester.ac.uk}
\author{Felix Pertl}
\author{Lubuna Shafeek}
\author{Scott R. Waitukaitis}
\affiliation{ 
$^1$Institute of Science and Technology Austria, Am Campus 1, 3400 Klosterneuburg, Austria
}

\date{\today}

\begin{abstract}
Scanning Kelvin probe microscopy (SKPM) is a powerful technique for macroscopic imaging of the electrostatic
potential above a surface. Though most often used to image work-function variations of conductive
surfaces, it can also be used to probe the surface charge on insulating surfaces. In both cases, relating the measured
potential to the underlying signal is non-trivial. Here, we derive general relationships between the measured SKPM
voltage and the underlying source, revealing either can be cast as a convolution with an appropriately
scaled point spread function (PSF). For charge that exists on a thin insulating layer above a conductor,
the PSF has the same shape as what would occur from a work-function variation alone,
differing by a simple scaling factor. We confirm this relationship by: (1) backing it out from
finite-element simulations of work-function and charge signals, and (2) experimentally comparing
the measured PSF from a small work-function target to that from a small charge spot. This scaling factor
is further validated by comparing SKPM charge measurements with Faraday cup measurements
for highly charged samples from contact-charging experiments. Our results highlight a hereto unappreciated
connection between SKPM voltage and charge signals, offering a rigorous recipe to extract either
from experimental data.
\\[10px]
\\[10px]
\textbf{This is a pre-print.  Please check for corrections/modifications when the article is published.}
\\Copyright \copyright~2025.
This manuscript version is made available under the CC-BY-NC-ND 4.0\\
license \url{http://creativecommons.org/licenses/by-nc-nd/4.0/}.
\end{abstract}

\maketitle

\section{Introduction}

Quantitative measurement of surface charge is important toward
understanding processes ranging from contact charging and
electrostatic discharge to lubrication and adsorption
\citep{Lacks2019Aug, Kumada2003May, Li2022Oct, Fang2024Mar}.
In turn, these processes find applications
in realms as diverse as industrial manufacturing, fabrication of nano-materials, or even cell adhesion
\citep{MendezHarper2024Jan, Hackl2022Nov,
Metwally2019Nov, Sobolev2022Nov}.
Scanning Kelvin probe microscopy (SKPM) is a technique for
measuring the electrostatic potential near surfaces \citep{Zisman1932Jul, Craig1970Feb, Lenton2024Apr}, and, although most often used to measure work functions of conducting surfaces\citep{Nazarov2012Oct, Nazarov2019Aug}, it can also give information
about the surface and bulk charge density of 
insulating samples \citep{Nalbach2000Apr, Martin2008Apr}. 

\begin{figure}[h!]
    \centering
    \includegraphics{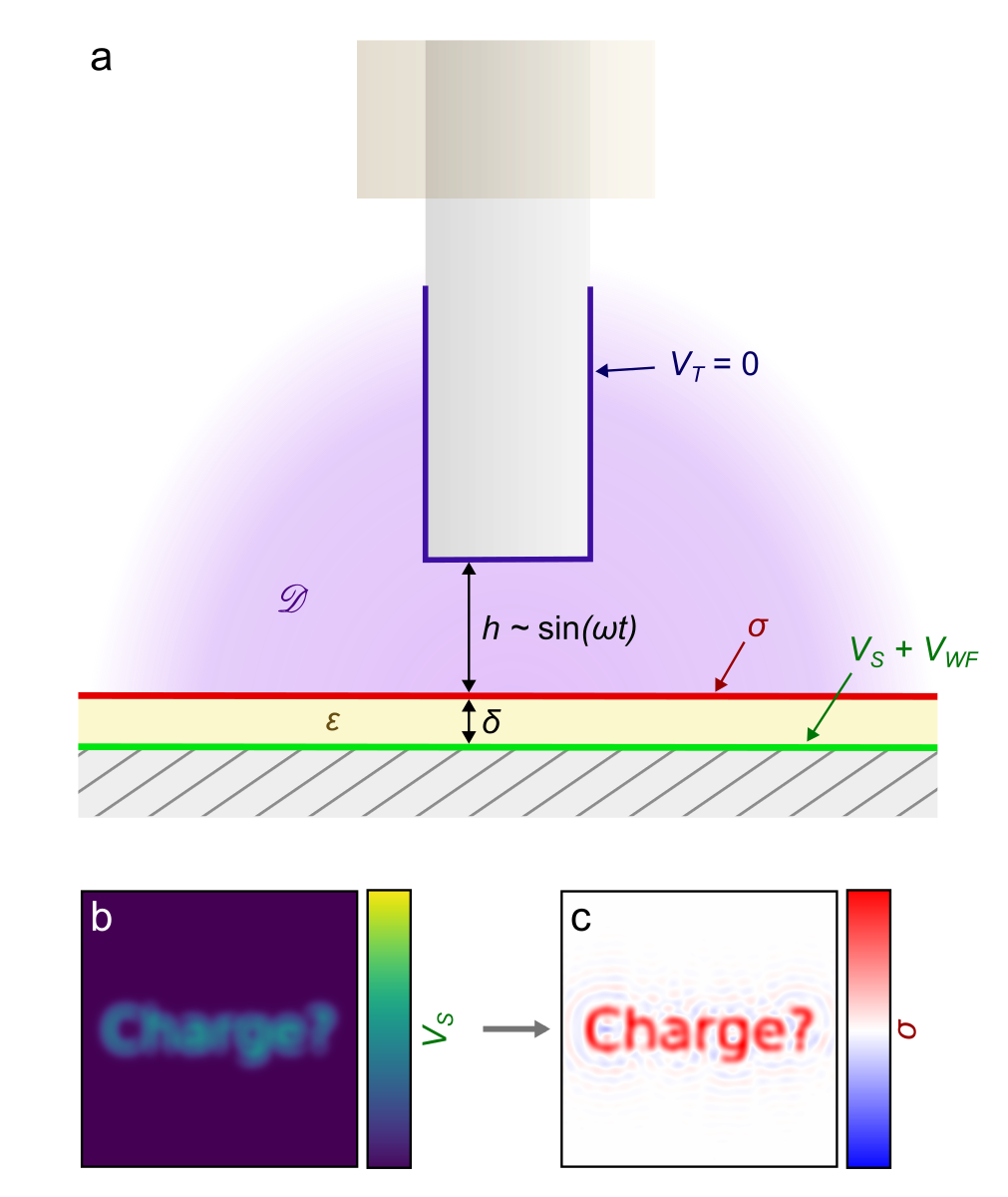}
    \caption{\textbf{Depiction of
    surface charge measurement with SKPM.}
    \textbf{a}~The probe-sample system forms a capacitor.
    When the probe is vibrated, the measured current depends
    on the geometry and properties of the system (the relevant parameters are discussed further in Section~\ref{sec:theory}).
    \textbf{b,c}~Converting the measured potential (\textbf{b})
    into an estimate for the surface potential (\textbf{c})
    requires accounting for the complexity of this system.}
    \label{fig:1}
\end{figure}

A key difficulty in using SKPM on charged surfaces is the quantitative conversion of the measured potential (with units of volts) into surface charge density (with units of coulombs per square meter). Figure~\ref{fig:1} illustrates the problem; the SKPM signal is related to the current induced in a probe as it is vibrated above a sample, which in turn depends on the electrostatic interaction between the probe/sample, the system geometry, and the electrostatic properties of the sample/medium. To recover an estimate for the surface charge, one must account for distortions in the measured surface potential caused by the geometry, as well as convert from units of potential to units of charge density.
While the electrostatic field profile due to a charge on a surface, and the connection between measured potential and charge is somewhat understood
from a theoretical point of view\citep{Sadeghi2013Jul,Somoza2020Feb}, the conversion from measured potential back to charge remains challenging.
One widely used heuristic is to assume a capacitor-like relationship between charge and measured potential, \textit{i.e.}~$\sigma=\tfrac{\delta}{\epsilon} V_S$, where $\sigma$ is surface charge density, $\delta$ is the insulator thickness, $\epsilon$ is its permittivity and $V_S$ is the measured voltage\citep{Heile2023Aug, Cai2016Jun, Barnes2016Jun, Palleau2010May}.  This expression is sufficient to give the correct units for charge. 
Though it is widely used, we have never seen a rigorous derivation of it.  Another shortcoming is that it does not account for the system geometry, \textit{i.e.}~the fact that the shape of the probe, its distance to the surface, and the thickness of the surface all `smear out' the underlying source.
Moreover, it is not clear when it is applicable. How thick can the insulator be before it fails?  How large can $\epsilon$ be? Hence, at present using this `capacitor heuristic' is at best an educated guess, as it lacks a rigorous framework to back it up.

In this work, we focus on rigorously understanding
the connection between signals from surface charge (SQ) and work function differences (WF) in SKPM. We begin by deriving a relationship between the
measured signal, SQ and WF. Both the signals from WF and SQ contribute cumulatively to the total measured signal: in both cases the measured signal can be formulated as a convolution of the underlying charge/work-function signal and an appropriately scaled point spread function (PSF). We show rigorously that  the PSF of a charge pattern on the surface of a thin dielectric above a conductive back electrode has the same shape as the PSF due to work function variations on the back electrode alone (\textit{i.e.,} absent the insulating layer), but indeed differs by the widely presumed scaling factor $\delta/\epsilon$. We directly test this by performing finite-element simulations corresponding to the two situations, which demonstrates how it fails if the insulator is too thick or the permittivity too high.  We further test it by comparing experimental measurements of both SQ and WF PSFs. Finally, we show how quantitative surface charge measurements can be acquired by using a PSF to deconvolve SKPM data, and validate measurements of total charge by comparing to results from a Faraday cup.

\section{Theory}\label{sec:theory}

SKPM involves scanning a vibrating conductive probe above a surface at a known frequency, and
measuring the current induced in the probe due to the local electrostatic
field at the same frequency using a lock-in amplifier \citep{Zisman1932Jul, Craig1970Feb, Lenton2024Apr}.
By additionally applying a DC voltage between the probe and the sample (or for thin insulating samples, a backing electrode below the sample) and using feedback, the lock-in current in the probe can be minimised. The SKPM potential image (Fig.~\ref{fig:1}\textbf{b}) corresponds to this voltage that minimises the lock-in current signal at different locations above the sample. In order to extract the surface charge (Fig.~\ref{fig:1}\textbf{c}), we need to account for all the factors which can contribute to the measured signal, such as system geometry, long-range electrostatic interactions between the probe and sample, and properties of the insulating layer.

For the geometry shown in Figure~\ref{fig:1}\textbf{a}, consisting of a thin homogeneous insulating layer above a conducting surface, we can take a first-principles approach to deriving the relationship between measured potential and the underlying source. Effectively, SKPM involves finding
\begin{equation}
    0 = i(t) \equiv \frac{dQ}{dt}
    \label{eq:current}
\end{equation}
where $i(t)$ is the current measured by the lock-in amplifier and
$Q(t)$ is the charge in the SKPM probe as a function of time $t$. Given the timescale of the probe oscillations is slow ($\sim$50 Hz), we can assume the system is electrostatic, and hence the charge distribution can be expressed as a surface integral involving the scalar potential $\phi$ on the probe surface, $\mathcal{S}_T$:
\begin{equation}
    Q = -\varepsilon_0 \oint_{\mathcal{S}_T}
        \nabla_x\phi(x)\cdot dn_x,
\end{equation}
where $\varepsilon_0$ is the vacuum permittivity,
$\nabla_x$ denotes the gradient operator with
respect to the spatial coordinate $x \in \mathbb{R}^3$,
and $dn_x$ is a unit area element
pointing normal to the surface.
If we consider a small time-dependent movement of
the tip, $dh$, we can rewrite Eq.~\ref{eq:current} as
\begin{equation}
    0 = \frac{dh}{dt}\frac{dQ}{dh}
        = \oint_{\mathcal{S}_T}
            \frac{d}{dh}\nabla_x\phi(x)\cdot dn_x.
    \label{eq:small-displacement}
\end{equation}

For an arbitrary domain, $\mathcal{D}$, bounded by a surface
$\mathcal{S}$, we can write the electrostatic potential
in terms of a Green's function, $G(x, \xi)$ for $(x,\xi) \in \mathbb{R}^3$,
\begin{equation}
\begin{split}
    \phi(x) = &\frac{1}{4\pi\varepsilon_0}\int_\mathcal{D} \rho(\xi) G(x, \xi)\ d\xi \\
        &+ \frac{1}{4\pi} \oint_\mathcal{S}
            V_\mathcal{S}\nabla_\xi G(x, \xi)\cdot dn_{\xi}
\end{split}
\end{equation}
where $\rho$ describes point sources (\emph{i.e.}, charges) within the domain
and $V_\mathcal{S}$ describes the potential on the boundaries.
The Green's function accounts for effects from system
geometry (\textit{i.e.}, the shape of the probe, the insulating layer,
and the probe location relative to the surface),
and $V_\mathcal{S}$ describes the potential on the boundaries.
We note that $G(x, \xi)$ vanishes at the boundaries
($\xi \in \mathcal{S}$).
For the system shown in Figure~\ref{fig:1},
choosing the tip potential as zero ($V_T = 0$), and the potential far
from the sample as zero ($V_{\xi \rightarrow \infty} = 0$),
leaves the only non-zero $V_\mathcal{S}$
terms corresponding to the integral over the sample
surface (\emph{i.e.}, the combination of applied potential $V_S$ and the
potential arising due to surface workfunction variations $V_{WF}$:
$V_S + V_{WF} \neq 0$).

If we limit our attention to a 2-D sample with
a 2-D surface charge distribution, $\sigma(\xi)$, on some plane, $\mathcal{S}_\sigma$, above but parallel to the lower boundary, $\mathcal{S}_\Phi$, the potential simplifies to two surface integrals
\begin{equation}
\begin{split}
    \phi(x) &= \frac{1}{4\pi\varepsilon_0}\oint_{\mathcal{S}_\sigma}
        \sigma(\xi) G(x, \xi)\ d\xi \\
    &\quad + \frac{1}{4\pi} \oint_{\mathcal{S}_\Phi}
        (V_{\text{WF}}(\xi) - V_\text{S})\nabla_\xi G(x, \xi)\cdot dn_{\xi}.
\end{split}
\end{equation}
Substituting this expression for the potential into
Eq.~\ref{eq:small-displacement} gives
\begin{equation}
\begin{split}
    0 = \oint_{\mathcal{S}_T}
            &\frac{d}{dh}\nabla_x
         \bigg( \frac{1}{\varepsilon_0} \oint_{\mathcal{S}_\sigma}
        \sigma(\xi) G(x, \xi)\ d\xi \\
    & + \oint_{\mathcal{S}_\Phi}
        (V_{\text{WF}}(\xi) - V_{\text{S}})\nabla_\xi G(x, \xi)\cdot dn_{\xi}
        \bigg) \cdot dn_x.
\end{split}
\end{equation}
By noting that the scalar potential is finite, we can
re-arrange for the applied surface voltage $V_S$:
\begin{equation}
\begin{split}
    V_\text{S} \oint &\nabla_{\xi} U(\xi)\cdot dn_\xi \\
    &= \frac{1}{\varepsilon_0} \oint \sigma(\xi) U(\xi)\ d\xi
        + \oint V_\text{WF}(\xi) \nabla_{\xi} U(\xi)\cdot dn_\xi.
\end{split}
\end{equation}
where we have dropped the domains in the above surface integrals as
they are implied by the integrand, and we have introduced
\begin{equation}
    U(\xi) = 
    \oint_{\mathcal{S}_T}
            \frac{d}{dh}\nabla_x G(x, \xi)\cdot dn_x.
\end{equation}

\begin{figure*}
    \centering
    \includegraphics{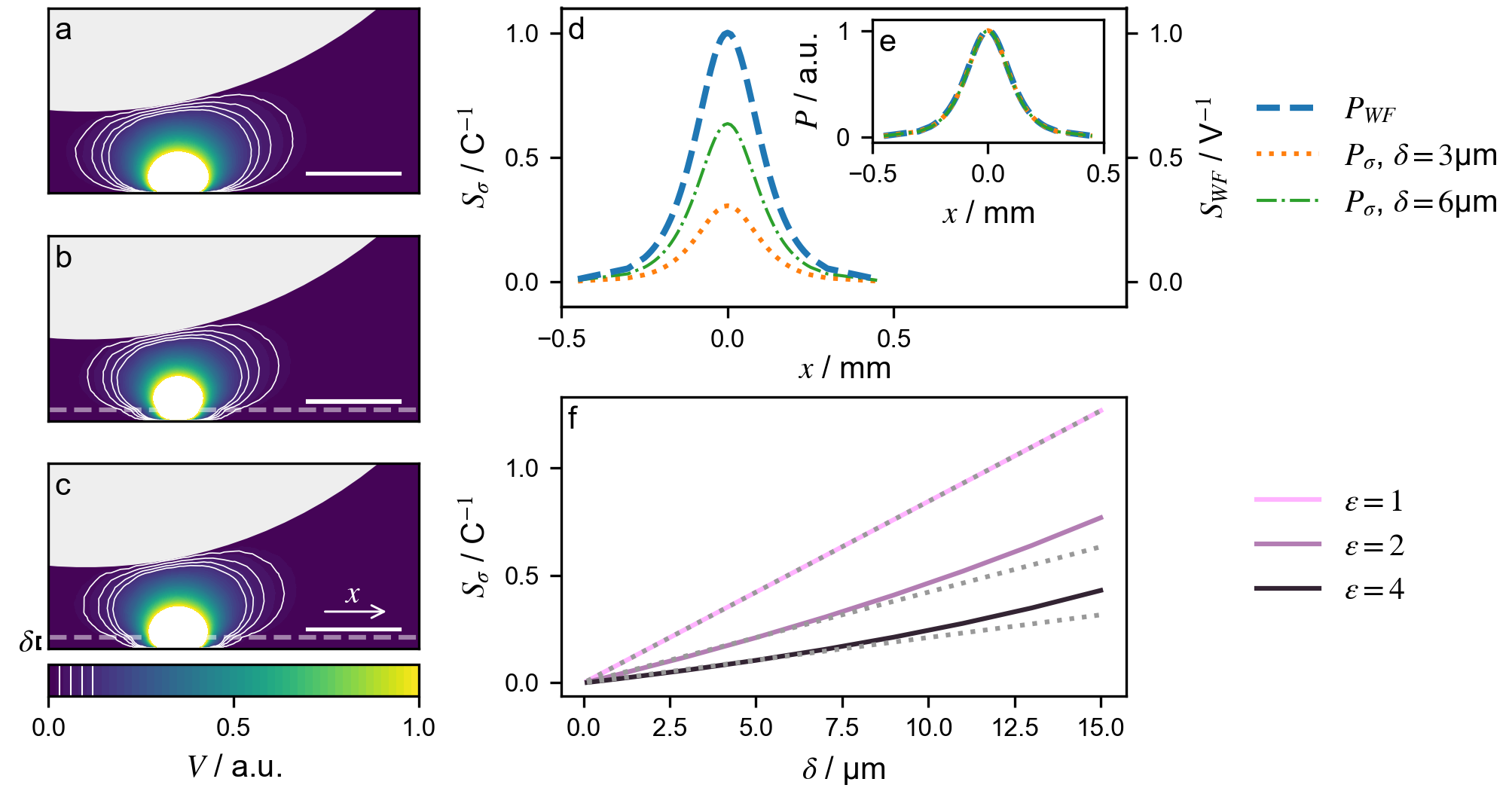}
    \caption{\textbf{Simulations showing the potential distribution
    from charge and work function point-like sources and the
    corresponding PSFs.}
    \textbf{a--c}~Normalised potential distributions between the
    probe (top hemisphere) and surface (bottom edge) for
    \textbf{a}~work function point with no insulator,
    \textbf{b}~charge point on insulator,
    and \textbf{c}~work function point beneath insulator.
    Scale bar shows 50~$\mu$m.  The dashed line marks the boundary
    of the insulator.  Contours (white lines) are equally spaced, illustrating
    the similarity in the potential gradient near the probe surface.
    \textbf{d}~Simulated PSFs for surface work
    function ($S_{WF}$) and charge ($S_\sigma$) points at two
    different heights above the surface
    \textbf{e}~PSFs in normalised units ($P$), emphasising
    the similarity in shape between simulated PSFs $S_{WF}$ and $S_\sigma$.
    \textbf{f}~PSF amplitude as a function of layer thickness ($\delta$) for different permittivities ($\epsilon$). For small insulating layer thicknesses, the relationship
    between work function and charge are linear, as illustrated by the dashed lines.}
    \label{fig:2}
\end{figure*}

To convert the above expression for the voltage at a single
point into a potential map describing the surface charge or
work function, we first consider how
the Green's function changes when the tip is translated.
For an infinitely wide uniform/flat sample, the Greens's function
is translationally invariant, \textit{i.e.}
$G(x, \xi - \eta) = G(x, \xi)$.
Substituting $\xi \rightarrow \eta - \xi$ into the above, we can
write the voltage distribution $V(\eta)$ for a scan over
the surface as
\begin{equation}
\begin{split}
    V_\text{S}(\eta) = \oint
        &\frac{u_0(\xi)}{u_2(\eta)}\sigma(\eta - \xi)\ d_\xi \\
    &+ \oint \frac{u_1(\xi)}{u_2(\eta)}V_{\text{WF}}(\eta - \xi)\ d_\xi
\end{split}
\label{eq:measurement-1layer}
\end{equation}
where we have introduced
\begin{eqnarray*}
    u_0(\eta) &\equiv& \frac{1}{\varepsilon_0} U(\eta), \quad \\
    u_1(\eta) &\equiv& \nabla U(\eta)\cdot n,\ \textrm{and}\quad \\
    u_2(\eta) &\equiv& \oint \nabla U(\eta)\cdot dn.
\end{eqnarray*}
The factors $P_{\sigma} = \frac{u_0}{u_2}$
and $P_{\text{WF}} = \frac{u_1}{u_2}$ correspond to the PSFs for
charge and work function in this geometry.

Calculating either of these PSFs analytically from first principles is difficult.  As we will show, however, both PSFs can be calculated in finite-element simulations and measured experimentally. One interesting outcome we can see analytically is the relationship between $u_0$
and $u_1$. If we determine either
of these, we can easily
calculate the other by simply integrating or taking the
derivative.
Furthermore, because in typical SKPM operation the distance between the probe and the sample is large
compared to the thickness of the insulating layer ($\delta$),
we can approximate the derivative with
\begin{equation}
    u_0(\eta; z = \delta) \approx \frac{\delta}{\varepsilon_0} u_1(\eta; z = 0)
\end{equation}
since $u_0(z = 0) = 0$ from the definition of our Green's
function.
Replacing the vacuum permittivity by the material
permittivity $\epsilon$ for
the region between the charge surface and backing electrode,
we recover the simple capacitor relationship between the measured
SKPM potential and charge density\citep{Burgi2004Nov, Cai2016Jun}
\begin{equation}
    \label{eq:capacitor}
    P_{\sigma} \approx \frac{\delta}{\varepsilon} P_{\text{WF}}.
\end{equation}
As we mentioned in the introduction, this ``capacitor heuristic'' is widely presumed throughout the literature, but to the best of our knowledge has not been derived.  Moreover, our analysis here shows that the relationship is deeper than simple proportionality of surface charge density and voltage. This is because it is not \textit{a priori} clear that the PSFs in the two cases should be the same; yet they are.  This equivalence has significant implications for calibration and measurement with SKPM: it suggests that measuring either the
surface charge or work function PSF can give us complete information about the other.

\section{Results \& Discussion}

\subsection{Numerical determination of PSF shapes}
As a first verification of the predictions from the previous section, and to explore over which length and permittivity scales they are valid (Eqs.~\ref{eq:capacitor}), we built a COMSOL model for our probe-sample system
(full details are given in Section~\ref{sec:methods}).
Figure~\ref{fig:2}\textbf{a--c} show the electric potential between
the a probe and the surface for the three relevant cases:
a point-like potential with no insulating layer,
a point-charge on an insulating layer,
and a point-like potential beneath an insulating layer.
These cases are for a relatively thick ($\delta = 6$ \textmu{m}) insulating layer
with a relatively high permittivity
($\varepsilon/\varepsilon_0 = 4$). The probe position and diameter
are comparable to a realistic SKPM measurement with diameter
500 \textmu{m} and probe-to-surface distance of 60 \textmu{m}.
Near the point-source, the shape of the potential differs
significantly, demonstrating how in general
$G_\varepsilon \neq G_{\varepsilon_0}$.
However, far from the point-source near the probe surface,
we can see that the resulting electric field is quite similar for the three cases, supporting the applicability of
Eq.~\ref{eq:capacitor}.

Accordingly, the behavior we predict occurs in the simulated PSFs
(Fig.~\ref{fig:2}\textbf{d}): all three PSFs have similar shapes, despite corresponding to slightly different physical situations.   When the simulated PSFs are normalised by the peak value
(Fig.~\ref{fig:2}\textbf{e}), we see this more clearly as they collapse to the same curve.  Figure~\ref{fig:2}\textbf{f} plots the ratio between
the work function PSF and the charge density PSF. The linear relationship given by Eq.~\ref{eq:capacitor} is valid at small $\delta$ (relative to the sample-probe separation) for all $\varepsilon$, but the range of this becomes shorter with increasing $\varepsilon$.  Hence, we can conclude that the ``capacitor heuristic'' is valid for typical insulators (\textit{e.g.}~those with permittivities $\epsilon/\epsilon_0\lesssim 5$) and typical probe geometries (\textit{i.e.}, large probe-to-sample distance compared to the insulator
thickness).  For substantially thicker surfaces or higher permittivities, it can indeed be expected to fail miserably.

\subsection{Experimental comparison of PSF shapes}
Beyond simulations, we can validate our ideas experimentally. We do this by fabricating PSF ``calibration targets'', \textit{i.e.}, very small source features with which we can scan our SKPM to effectively obtain PSFs. These are a small disc of metal and a point-like charge spot for the WF and SQ cases, respectively. For the metal disc, we performed lift-off
lithography to evaporate a small gold disc on a silicon wafer
and then deconvolved the measured signal by the disc shape to
estimate the PSF \citep{Lenton2024Apr}.
For the charge target, we deposited a localized spot (2 \textmu{m} diameter) of charge on a wafer with a thermally grown SiO$_2$ layer (thickness $\delta=3$ \textmu{m}) using a plasma focused ion beam (PFIB). We confirmed that the surface geometry wasn't changed significantly as a result of the PFIB process by imaging identically prepared samples with atomic force microscopy.  Full details are provided in Section~\ref{sec:methods}.

\begin{figure}
    \centering
    \includegraphics{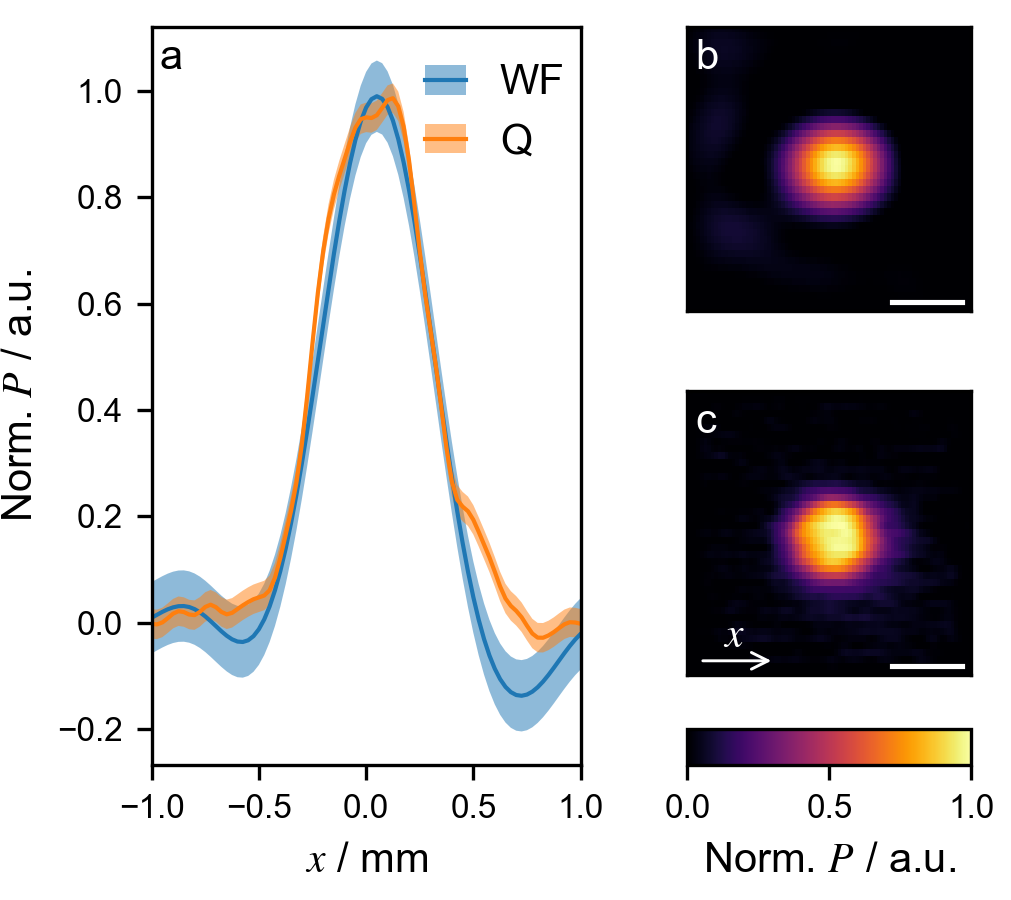}
    \caption{\textbf{Experimentally measured PSFs for a work function (WF) and charge (Q)
    signal.}
    \textbf{a}~Comparison between slices of the PSF calculated
    using the charge target and work function target.
    Shaded regions show $2\sigma$ uncertainty estimate based on the
    measurement signal-to-noise ratio.
    \textbf{b}~The estimated PSF for a work function target.
    \textbf{c}~The estimated PSF for a charge target.
    Scale bars indicate 500~$\mu$m.}
    \label{fig:3}
\end{figure}

SKPM scans for the two cases are shown in  Figure~\ref{fig:3}. We confirm again that both PSFs have essentially the same shape; normalizing each by the peak value, they largely collapse on top of each other. The one notable difference is the relative size of the error bars. The charge spot produced a significantly larger SKPM signal, much larger than the SKPM system noise. The stronger charge signal also allows measurement of higher frequency parts of the PSF signal, revealing broadening of the PSF near the peak, as we saw in our previous work with slower scan PSFs \citep{Lenton2024Apr}. The strong negative region in the WF signal around $x=0.5$ is likely an artifact resulting from a combination of the weak signal and absence of high frequency information. In the SI, we show the power spectra for these signals, which more clearly demonstrate the improved signal-to-noise achievable with the charge target. Hence, once again we see that our main result is validated.

\subsection{Surface charge measurement with SKPM}
To measure the surface charge with SKPM, we need to separate the
contributions from the material/geometry from the contributions
due to surface charge and then convert the potential into charge.
For a relative charge measurement, \emph{e.g.}, charge added due to
a contact with another material, we can simply scan the surface
before and after the charging event and compare the potentials
\begin{equation}
    V_{s,rel} = V_{s,2} - V_{s,1} =
        \oint \frac{u_0}{u_2}(\sigma_2 - \sigma_1)
\end{equation}
where $\sigma_{1,2}$ are the surface charge densities before/after
the charging event and the surface integral corresponds to the first
term in Eq.~\ref{eq:measurement-1layer}.
To estimate the relative charge, we can deconvolve the measured
potential by the relevant PSF\citep{Lenton2024Apr}
\begin{equation}
    \sigma = P_\sigma^{-1} \otimes V_{s,rel}
\end{equation}
where we have used $P_\sigma^{-1} \otimes$ to denote the
deconvolution.
This approach is sufficient for many types of contact charging
experiments.

\begin{figure}
    \centering
    \includegraphics{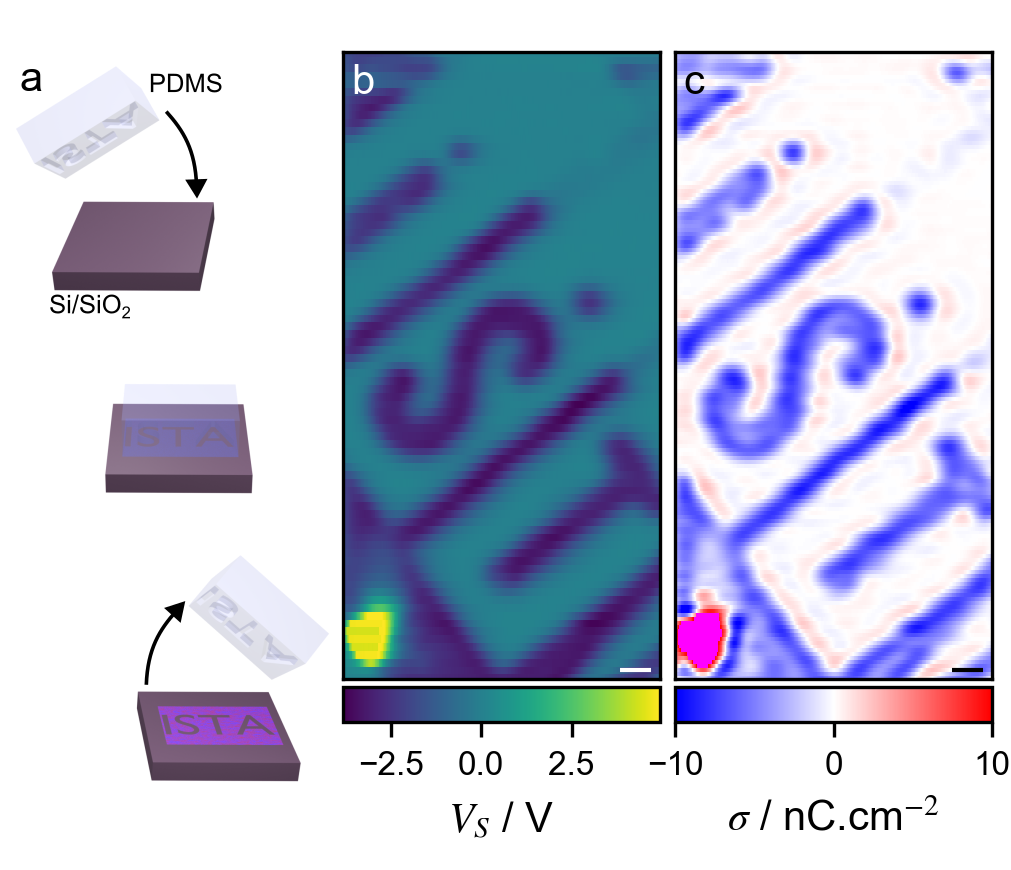}
    \caption{\textbf{Example SKPM potential measurement and the deconvolution
    to give the surface charge density.}
    \textbf{a}~A charge pattern is created by pressing a PDMS stamp against a
    silicon wafer with a thin oxide layer.
    \textbf{b}~The resulting SKPM
    signal~\textbf{b} can be deconvolved with the PSF to give the surface
    charge density~\textbf{c}.
    Scale bars show 1~mm.}
    \label{fig:4}
\end{figure}

Figure~\ref{fig:4} depicts a typical contact charging experiment:
a sample of PDMS is contacted with a silicon wafer with a thin
insulating layer of SiO$_2$ (thickness $\delta=3$ \textmu{m}).
After removing the PDMS the surface becomes
charged\citep{Baytekin2011Jul, Sobolev2022Nov, Pertl2022Dec}.
Figure~\ref{fig:4}\textbf{b} shows the relative change in potential
before/after the contact event.
By deconvolving the signal by the PSF, we recover the surface
charge left by the charging event (Fig.~\ref{fig:4}\textbf{c}).
We see that the charging is non-uniform, largely due to the
pattern we used to cure the PDMS (our institute logo),
and partially due to the peeling direction when removing the PDMS stamp,
as has been previously described\citep{Sobolev2022Nov, Reiter2021Aug}.
One region acquired a significant positive charge,
indicated by the pink region in Fig.~\ref{fig:4}\textbf{c}
where we were unable to measure the charge accurately
due to the limited voltage range of our SKPM.

As further validation for the scaling factor used to convert the
work function PSF into a charge PSF, we compared SKPM-derived charge measurements to those made with an independent charge-measurement technique.
We placed a silicon wafer with a silicon dioxide layer
inside a Faraday cup, contacted the wafer
with a PDMS stamp and measured the total charge remaining on the
wafer after the PDMS was removed from the Faraday cup.
We then transferred the wafer to the SKPM, measured the surface charge density and integrated to calculate the net charge.
\begin{figure}
    \centering
    \includegraphics{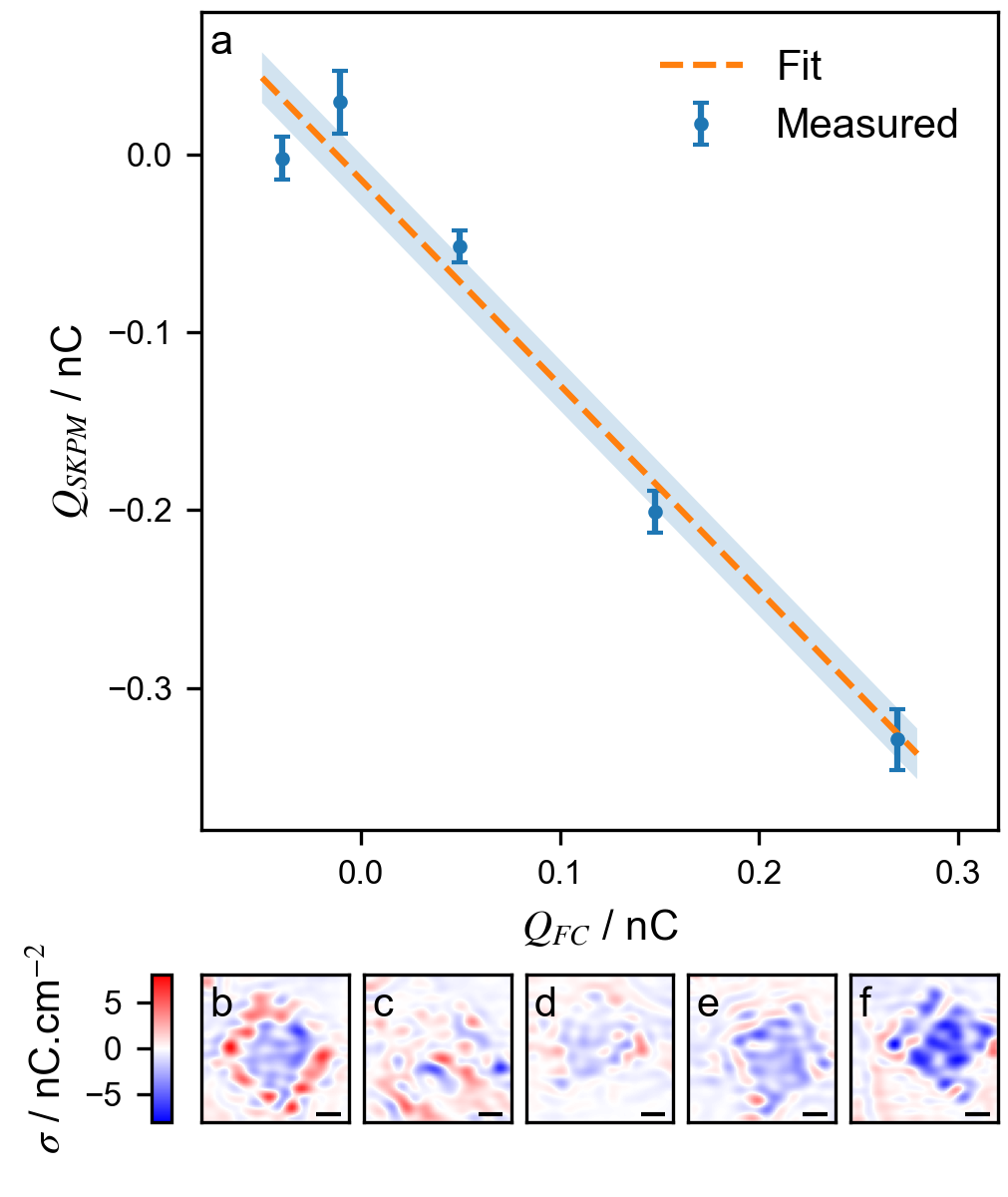}
    \caption{\textbf{Comparison between SKPM and Faraday cup charge measurements.}
    \textbf{a}~Plot showing SKPM $Q_{SKPM}$ and Faraday Cup $Q_{FC}$ charge measurements.
    \textbf{b--f}~show the SKPM scans for samples in order of increasing charge for
    each data point in the plot.
    Scale bars show 1~mm.  Error bars show 1$\sigma$ uncertainty in measurement (see Section~\ref{sec:methods} for details).  Shaded region shows uncertainty in fit.}
    \label{fig:5}
\end{figure}
The results (Figure 5) show good agreement, especially in the correlation factor: -1.15 $\pm$ 0.14~C/C.
The sign difference corresponds to a difference in
convention used by the electrometer and SKPM.
The main sources of error, including the DC shift,
was estimation of the background
charge level, including how much charge was already on
the surface before contact, and charge transferred to
the sample during handling.
For relative charge measurements, we find that the accuracy is comparable
to the Faraday cup charge estimation (for these samples).
Other factors contributing to the difference in slope may
be related to variations in oxide layer
thickness, variations
in permittivity, or the limitations of
the simple capacitor-like approximation we used for the PSFs.

\section{Conclusion \& Perspectives}
We have derived a relationship between the PSFs for work function and surface charge density in SKPM,
and validated it in both in simulations and experimentally.
We found that, at least for thin insulating layers and low dielectric constants, the widely-presumed capacitor scaling is valid.  However, we furthermore have shown that the PSFs necessary to deconvolve SKPM data for the charge/potential case have the same shape, differing only by the capacitive scaling factor.  While we have focused on the macroscopic SKPM technique, our results should also be extendable to the nanoscopic counterpart of Kelvin Probe Force Microscopy (KPFM), although the differences in length scales may
limit the applicability. Furthermore, we validated that SKPM is a suitable tool for quantitative measurement of surface charge on thin insulating materials. Importantly, measurements can be
calibrated without requiring separate simulations or calibrations
for the individual charge or work function cases. This has two important consequences; first, it enables the fabrication of calibration targets with much higher signals compared to conventional work function-based targets, overcoming problems we previously identified with the fabrication of calibration targets for SKPM\citep{Lenton2024Apr}, and second, it removes the need for repeated calibration of different insulation-thickness combinations. 

\section{Experimental Section}
\label{sec:methods}

\paragraph*{Numerical modelling}
To simulate the work function and charge density PSFs we created
a electrostatics model in COMSOL Multiphysics 5.3a.
For calculating the work function PSFs, we positioned a small
potential disc on a grounded plate at different positions bellow
the SKPM probe.
To explore the effect of adding a thin insulating layer, we repeated
this simulation with a thin dielectric layer covering
the disc plate.
For calculating the charge PSFs, we added a dielectric layer above
the plate and added a point charge on the interface between the
dielectric layer and air.
To calculate the SKPM potential, we compared the current density
in the probe at two different heights, as we described
in our earlier work\citep{Lenton2024Apr}.

\paragraph*{Silicon dioxide wafer preparation}
Silicon dioxide wafer pieces with varying oxide thicknesses were prepared
by first cleaning using a two step solvent clean (Acetone \& Isopropyl alcohol
in an ultrasonic bath) and then baking at grater than 120$^\circ$C (typically
300$^\circ$C) for more than 30 minutes.
Different cleaning conditions give different initial surface potentials.
We found the baking step was important for achieving a uniform initial charge
distribution.

\paragraph*{PFIB target fabrication}
To fabricate the point-like charge target, we used a Helios G4 PFIB UXe
with e-beam and ion-beam to create a small charge spot on a silicon dioxide wafer.
Clean 3~$\mu$m oxide wafers were mounted on SEM pin stubs using
double sided carbon tape.
Samples were loaded into the PFIB and the system was pumped to vacuum.
After waking the e-beam and ion-beam, we positioned the e-beam and
ion-beam at the corner of the wafer in order to link the stage and
orientate the sample.
The ion-beam was set to 10~pA current and 30~kV acceleration voltage.
Both the ion- and e-beam imaging were then stopped, this was important
to prevent unwanted charging of the sample due to scanning of the
beams in imaging mode.
The sample was then orientated perpendicular to the ion-beam and
positioned so the ion beam was focused directly above the
desired target location.
The ion-beam was configured to write a circular pattern using the
default milling parameters for Si.  
We found that writing a circle with diameter 2~$\mu$m,
with dosage per exposure of 1.0$\times$10$^{-11}$~pC/$\mu$m$^2$ and
1~$\mu$s dwell time produced a target that could be measured
without saturating the SKPM.
Total exposure time was approximately 4 seconds with an
accumulated dosage of approximately 7~pC/$\mu$m$^2$
(total dosage of approximately 21~pC).
The system was then shutdown, the chamber vented and the sample
removed.
The sample was transferred immediately to the SKPM where argon
was used to reduce the humidity above the sample in order
to reduce charge decay\citep{Navarro-Rodriguez2023Feb}.

\paragraph*{PDMS stamp fabrication}
For creating charge patterns we fabricated PDMS stamps using
a traditional soft-lithography approach.
New silicon wafers (p-type, Boron, $\langle$100$\rangle$, polished) 
were cleaned using isopropanol
in an ultrasonic bath for 2 minutes, dried with nitrogen,
and baked at 110\degree C for 2 minutes to remove
any residual water or isopropanol.
SU-8 (GM1075, Engineering Solutions) was spin coated on the wafer at 500~rpm
for 15~s to spread
the SU-8 and then 100~s at 950~rpm to achieve the desired
thickness ($\sim$200~$\mu$m).
The wafer was baked for 10 minutes at 120\degree C.
The wafer was loaded into the mask aligner (EVG 610 with 
mercury bulb light source, EV Group)
and exposed to 750~mJ/cm$^2$
with the target pattern.
Patterns were designed using Creo Parametric and printed on 0.18~mm
Polyethylene terephthalate (PET) film (JD PhotoData).
The mask was mounted on a 1.5~mm glass plate to more easily load
it into the mask aligner.

Post-exposure, the wafer was baked at 95\degree C for 1 hour and then
developed in SU-8 developer with mild agitation,
before being rinsed with isopropanol and hard baked for 5 minutes at 135\degree C.
To allow easy removal of the PDMS stamp from the SU-8 stamp,
the wafer was dry silinised:
samples to be silinised were placed in a vacuum desiccator with a small
quantity of silane (448931-10G, Sigma-Aldrich),
pressure was reduced, vacuum turned off, and left for 2-3 hours.
For the PDMS stamps, we prepared 50~g of PDMS (SYLGARD 184, Dow Chemical Company)
with a 10:1 ratio (50~g is sufficient to cover three 100~mm wafers).
PDMS was poured over the SU-8 stamps and degassed using a vacuum desiccator and
nitrogen gun to remove any bubbles before curing in an oven for 4 hours at 80\degree C.
PDMS stamps could then be removed from the wafer when required.

\paragraph*{Faraday cup charge measurements}
To verify the SKPM charge measurement, 10$\times$10~mm
Si wafer pieces with 3~$\mu$m thermally grown oxide were cleaned, discharged, and
mounted in a custom made Faraday cup connected to an
electrometer (B2987B, Keysight).
Prior to measurement, the electrometer was self-calibrated according to the manual
and left for approximately 1 hour to allow it to stabilise.
The Faraday cup was purged with clean dry air prior to loading each sample, excess
humidity in the Faraday cup would often cause large drifts in the electrometer signal.
The electrometer was zeroed and a piece of PDMS
(approximately 3$\times$3~mm) attached to a
wooden stick was contacted with the wafer sample.
The electrometer value was recorded after the PDMS piece was removed
from the Faraday cup.
After contact, samples were transferred to the SKPM to be scanned.

\paragraph*{SKPM charge estimation}
We scanned the wafer with SKPM and integrated the surface charge
density to give the net charge.
Due to how the sample was mounted, it is difficult to scan the whole wafer, so only
the central region around where the PDMS was contacted with the wafer was scanned.
To estimate the error, we calculated the standard deviation along the
edge of the scanned region to give an estimate for the variation in background signal
across the wafer.

\section*{Acknowledgements}
This project has received funding from the European
Research Council (ERC) under the European Union’s
Horizon 2020 research and innovation programme (Grant
agreement No. 949120). This research was supported by
the Scientific Service Units of The Institute of Science
and Technology Austria (ISTA) through resources provided by the
Miba Machine Shop, Nanofabrication
Facility, Scientific Computing Facility,
and Lab Support Facility.
The authors wish to thank Dmytro Rak and
Juan Carlos Sobarzo for letting us use their equipment.
The authors wish to thank Evgeniia Volobueva for advice in
preparing PFIB samples.
The authors wish to thank the contributions of the whole
Waitukaitis group for useful discussions and feedback.

\section*{Author Declarations}
\subsection*{Conflict of Interest}
The authors have no conflicts to disclose.
\subsection*{CRediT Author Statement}
IL: Conceptualization; Formal analysis; Simulation; Investigation;
Writing -- Original Draft. FP: Investigation; Resources.
LS: Resources. SW: Writing -- Review \& Editing; Supervision; Funding acquisition.

\section*{Supplementary Information}
Additional supporting figures can be found in this article's
supplementary material.

\section*{References}

\bibliography{main}

\section*{Table of Contents Text}
Scanning Kelvin probe microscopy (SKPM) is a powerful technique for macroscopic imaging of the electrostatic potential above a surface, providing insight into sample work function and charge variations.
General relationships are derived connecting the measured SKPM voltage image and
the underlying work function or charge pattern.  These results are confirmed both numerically and experimentally.

\end{document}


\maketitle

\newcommand{\degree}{$^\circ$}
\renewcommand{\thepage}{S\arabic{page}}
\renewcommand{\thesection}{S\arabic{section}}
\renewcommand{\thetable}{S\arabic{table}}
\renewcommand{\thefigure}{S\arabic{figure}}

\section{Atomic force microscopy scans of PFIB samples}
\begin{figure}
    \centering
    \includegraphics[width=0.5\linewidth]{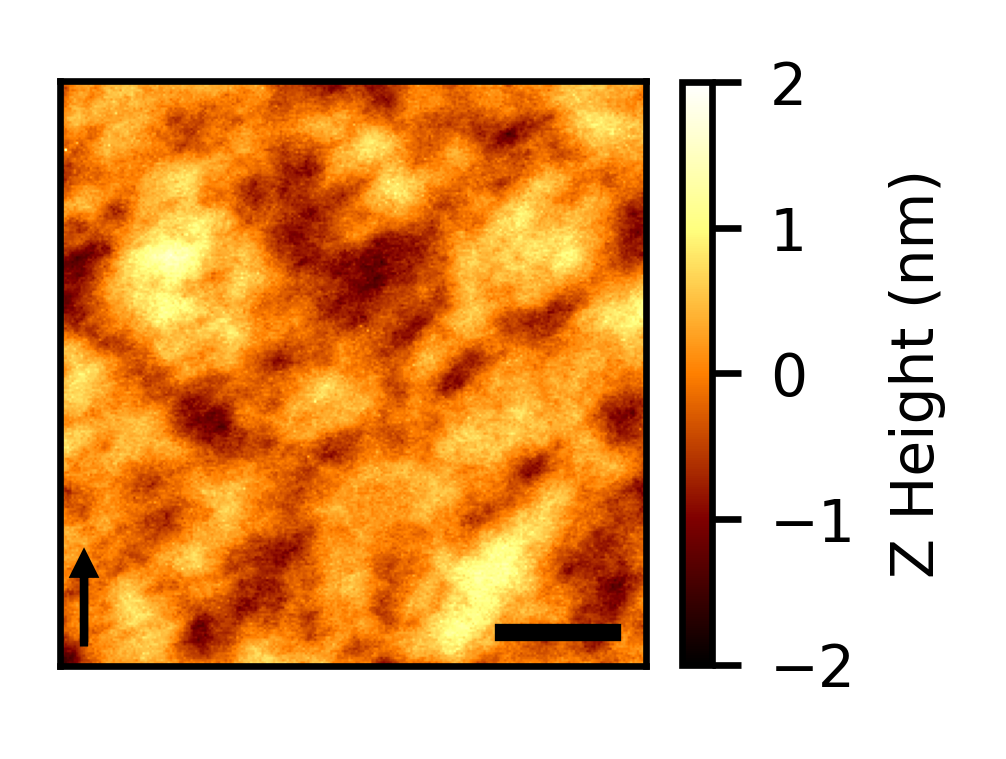}
    \caption{AFM scan of a wafer with $3\mu$m
    oxide layer after PFIB processing.  No
    obvious signs of significant surface damage are present
    with these PFIB settings. Scalebar is 5 $\mu m$. Arrow indicates scan direction in the secondary scan axis.}
    \label{fig:si-1}
\end{figure}

To verify that the surface geometry is not changed
significantly after PFIB, we performed over 400 scans with an atomic force microscopy (AFM) to check the surface for signs of damage. 
Figure~\ref{fig:si-1} shows one AFM surface profile of a sample prepared using the same PFIB settings as the surface charge target. No significant alteration to the surface geometry after PFIB was observed.

\section{Noise analysis of PFIB scans}
\begin{figure}
    \centering
    \includegraphics[width=\linewidth]{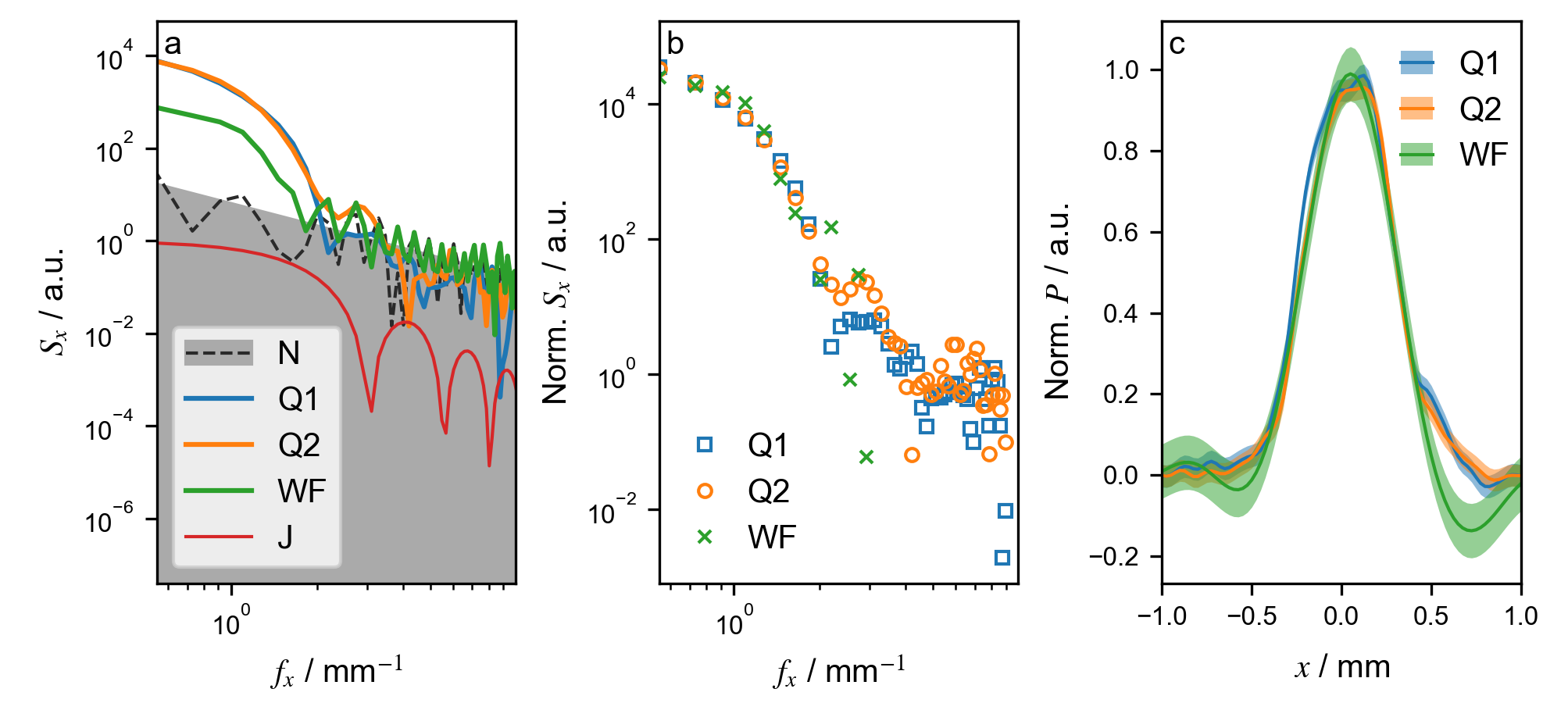}
    \caption{\textbf{a}~Power spectral density plots of
    different SKPM scans.  \textbf{b}~Normalised
    power spectra density plots for the calibration
    target scans.  \textbf{c}~Estimated PSFs.}
    \label{fig:si-2}
\end{figure}

In addition to the PSF measurement shown in Figure~\ref{fig:3}, additional measurements were collected to compare the charge PSF and WF PSF estimates.
Figure~\ref{fig:si-2}\textbf{a} shows the power spectral
density of several SKPM scans.
Two sequential charge PSFs were collected (Q1 \& Q2),
confirming no significant decrease in signal was
observed over the scan duration.
Additionally, a scan of a region without any charge or
work function features was collected (N) providing an
estimate for the measurement noise.
The signal from the two charge scans is much stronger
than the signal from the work function scan (WF), and
also extends out to higher frequencies before being
lost within the measurement noise.
For comparison, the Jinc used for the WF calibration
target correction is also shown (J) -- 
the bandwidth of the Jinc is higher than
the measurement bandwidth, confirming the WF calibration
target is sufficiently small for this measurement.

Figure~\ref{fig:si-2}\textbf{b} shows a comparison of
the power spectra for the three calibration target
measurements after normalisation.
Points below the noise estimate have been excluded.
All three PSF power spectra show a similar bandwidth.
There is some variation at higher frequencies, and
additional measurements or averaging may help improve
the PSF estimation.
Figure~\ref{fig:si-2}\textbf{c} shows the resulting
PSF estimates.